\documentclass[letterpaper,twocolumn,fleqn]{article} 

\usepackage{ist}
\usepackage{amsmath}
\usepackage{nccmath}
\usepackage{multicol}
\usepackage{multirow}
\usepackage{makecell}
\usepackage{array}
\usepackage{url}

\title{Validation of image systems simulation technology \\using a Cornell Box}

\author{Zheng Lyu, Krithin Kripakaran, Max Furth, Eric Tang, Brian Wandell, and Joyce Farrell \\ \\
Stanford Center for Image Systems Engineering, Stanford, California, 94305, USA \\
}

\date{
\today
}

\begin{document} 

\maketitle 



\begin{abstract}
We describe and experimentally validate an end-to-end simulation of a digital camera. The simulation models the spectral radiance of 3D-scenes, formation of the spectral irradiance  by multi-element optics, and conversion of the irradiance to digital values by the image sensor. We quantify the accuracy of the simulation by comparing real and simulated images of a precisely constructed, three-dimensional high dynamic range test scene. Validated end-to-end software simulation of a digital camera can accelerate innovation by reducing many of the time-consuming and expensive steps in designing, building and evaluating image systems. 
\end{abstract}

\section{Key Words}
\textit{Camera design, imaging system design, end-to-end simulation, computer graphics}

\section{Introduction}
\label{sec:intro}
Together with our colleagues, we are developing a programming environment for image systems simulation.  This work, initiated in 2003, was intended to simulate the optics and sensors in digital camera systems \cite{Farrell2012-ma, Farrell2003-rb}. The original simulations were validated by comparing simulated and real digital camera images using calibrated 2D test charts \cite{Farrell2008-sc, Chen2009-bj}. 

In recent years, we added physically-based quantitative graphics methods into the image systems simulation \cite{Liu2019-eo}. This enables us to calculate the spectral irradiance image of relatively complex, high-dynamic range (HDR) 3D scenes. We have used the methods to design and evaluate imaging systems for a range of applications, including AR/VR \cite{Lian2018-kc}, underwater imaging \cite{Blasinski2017-lr}, autonomous driving \cite{Blasinski2018-em, Liu2019-ds}, fluorescence imaging \cite{Farrell2020-tb, Lyu2021-ae} and models of the human image formation \cite{Lian2019-xr}. In each case, we carried out certain system validations - typically by comparing some aspect of the simulation with real data. The work we describe here is a further validation of the image systems simulations. In this case, we construct a Cornell Box, a 3D scene designed to demonstrate physically based ray tracing methods, and we compare predicted and measured camera data. 

The Cornell box (Figure \ref{Fig1}) was designed to include significant features found in complex natural scenes, such as surface-to-surface inter-reflections and high dynamic range \cite{Goral1984-wd}. The box contains one small and one large rectangular object placed near the side walls of the box. The box is illuminated through a hole at the top, so that some interior positions are illuminated directly and others are in shadow. The walls of the box are covered with red and green paper, providing colorful indirect illumination of the rectangular objects. In its original use, graphics software generated realistic renderings that were visually compared to the box using photographs \cite{Cohen1986-ld}, human perception \cite{Meyer1986-pb}, and CCD camera images \cite{Sumanta-qr}.  The papers showed that the graphics algorithms captured the main effects of the lighting and inter-reflections.  

We extend this approach in several ways.  First, we introduce into the simulation a camera with its optics and sensor (Google Pixel 4a).  The camera simulation approximates the camera optics and models many properties of the sensor, including pixel size, spectral quantum efficiency, electrical noise, and color filter array. An end-to-end simulation - from the Cornell box to the sensor digital values - enables us to go beyond visual comparisons to quantify the differences between the simulated and measured camera sensor data.  The ability to predict the numerical sensor data is particularly relevant for users who are designing image processing algorithms to render the images or using the sensor data for machine learning applications.

\section{Simulation pipeline}
An imaging system contains multiple components: a scene with light sources and assets, optical elements, sensors, image processing algorithms and displays. Our validation focuses on the first three elements: the scene, optics and sensor with the goal of predicting the unprocessed sensor data. We focus on matching the sensor data  because in typical use cases the designer has control over how the sensor data are analyzed.

The simulation workflow consists of three main components: 1) create a three-dimensional model to calculate the scene spectral radiance, 2) calculate the spectral irradiance image at the sensor, and 3) calculate the sensor image data.  In this section we describe the computational software tools that we use to achieve each of these tasks. 

\subsection{The Cornell box}
The first step in our simulations is to represent the 3D geometry of the Cornell Box and its contents, as well as the positions of the light sources and camera. This can be accomplished using several available computer graphics programs for modeling 3D objects and scenes.  For this project, we use Cinema 4D to create a 3D model of the objects and to represent the position of the light and the camera. The 3D coordinates, mesh, texture maps and other object information were exported as a set of text files that can be read by PBRT \cite{Pharr2016-yb}.

\begin{figure}[!ht]
  \includegraphics[width=1\columnwidth]{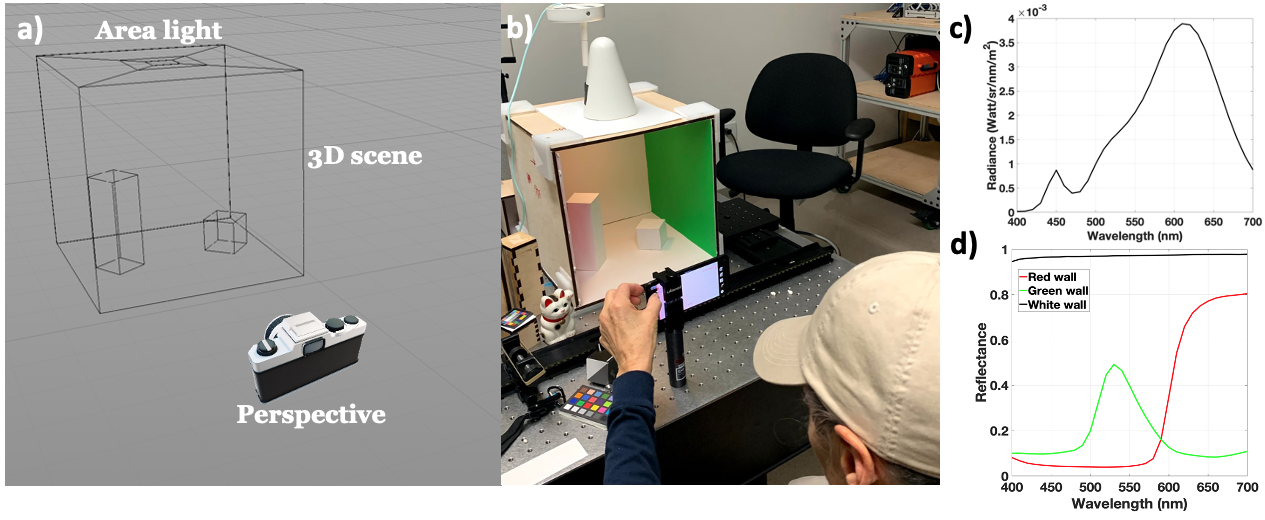}
  \caption{Virtual and Real Cornell box. a) Assets size, position of light source and camera perspectives are defined in Cinema 4D; b) Construction of a Cornell Box. c) Spectral power distribution of light source; d) Spectral reflectance of the red and green surfaces.}
  \label{Fig1}
\end{figure}

The real Cornell box and the objects within the box were built with wood, cut to the size specified by the model.  The surfaces of the inside walls and the rectangular objects were painted with white matte paint. The right and left walls were covered with green and red matte paper, respectively.  A hole was cut in the top of the Cornell Box and covered with a  diffuser; a light was placed on top of the diffuser. The spectral energy of the illuminant and spectral radiance of all surfaces within the Cornell box were measured with a PR670 spectroradiometer (See Figure \ref{Fig1}cd).

We used an open-source and freely available Matlab toolbox ISET3d \cite{Liu2019-eo, Blasinski2018-em} to read the PBRT files exported from Cinema 4d and to programmatically control critical scene properties, including the object positions, surface reflectance, and illuminant spectral energy. This spectral information is part of the material and light descriptions that are used by the rendering software, PBRT \cite{Pharr2016-yb}. We modeled the surfaces as matte materials; the light was modeled as an area light whose size matched the hole in the box made for the light.

\subsection{Optics modeling}
PBRT calculates how light rays propagate from the light source to surfaces in the scene, including the effects of surface inter-reflections. PBRT also traces light rays through the optical elements of a lens comprising multiple spherical or biconvex surfaces. The specific optical surfaces in the Google Pixel 4a, however, are unknown to us. Thus, instead of using PBRT to model the lens, we used an approximation.  Specifically, we modeled a diffraction-limited lens with (1) the same f-number and focal length, (2) a small amount of depth-dependent defocus, and (3) empirically estimated lens shading. 

We introduced the defocus by modeling the wavefront emerging from the diffraction-limited lens using Zernicke polynomial basis functions. A diffraction-limited lens has polynomial weights of 1 followed by all zeros. We slightly blurred the image by setting the defocus term (ANSI standard, j=4) to a positive value \cite{wiki:Zernike_polynomials}. We estimated the lens shading by measuring a uniform image produced by an integrating sphere and fitting a low order polynomial through the data.

\begin{figure}[!hb]
  \includegraphics[width=1\columnwidth]{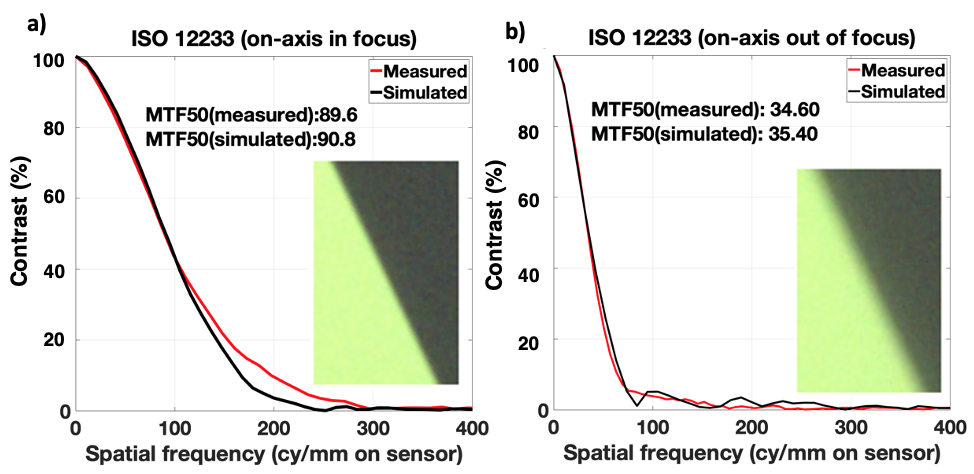}
  \caption{Modulation transfer function (MTF) comparison. a) On-axis in-focus MTF curves derived from images of the slanted bar. b) On-axis MTF curves when lens is focused 20 cm in front of the slanted bar.}
  \label{Fig2}
\end{figure}

In summary, we use PBRT to calculate the scene radiance.  We apply the lens blur and lens shading to calculate the irradiance at the sensor. 

We compared the spatial resolution of the simulated and measured data by calculating the modulation transfer function (MTF).  To calculate the MTF, we simulated a scene of a slanted edge target illuminated in the Cornell Box, and we captured Google Pixel 4a camera images of a real slanted edge target. We used the ISO 12233 method to calculate MTFs based on an ISO12233 analysis of the simulated  and real camera images of the slanted edge target. 
 
We made the MTF comparison for targets at two different distances. To model the defocus at different depths and field heights, we set the simulated defocus of a diffraction-limited lens.  Figure \ref{Fig2} compares predicted and measured MTF curves when the target is placed 50 cm away from the camera and is in focus (Figure \ref{Fig2}a), and  when the target is in the same position  but the camera is focussed 30 cm away (Figure \ref{Fig2}b).   The value of the Zernike polynomial defocus term was 1.225 um for the in-focus plane and 3.5 um for the out-of-focus plane. Adjusting the defocus alone brought the measured MTF and simulation into good agreement.

\subsection{Sensor modelling}
The simulated sensor irradiance image is converted to a sensor response using ISETCam. 

The Google Pixel 4a has a Sony IMX363 sensor, and many of the sensor parameters can be obtained from published sensor specifications (e.g., the pixel size fill factor, sensor size).  We made empirical calibration measurements in our laboratory to verify a subset of the published parameter values, specifically the spectral quantum efficiency, analog gain, and various electronic noise sources (see Table~\ref{tab:sensor} in the Appendix). We made these measurements using calibration equipment (PhotoResearch PR670 spectroradiometer) and OpenCam, a free and opens-source software program that controls camera gain and exposure duration, and retrieves lightly processed (“raw”) sensor digital values.

For example, to estimate the system sensor quantum efficiency, we captured Google Pixel 4a camera images of a Macbeth ColorChecker (MCC) illuminated by three different spectral lights available in a Gretag Macbeth Light Box. We measured the spectral radiance of each of the 24 color patches in the MCC and calculated the average R, G and B pixel values for the corresponding color patches.

\begin{figure}[!hb]
  \includegraphics[width=1\columnwidth]{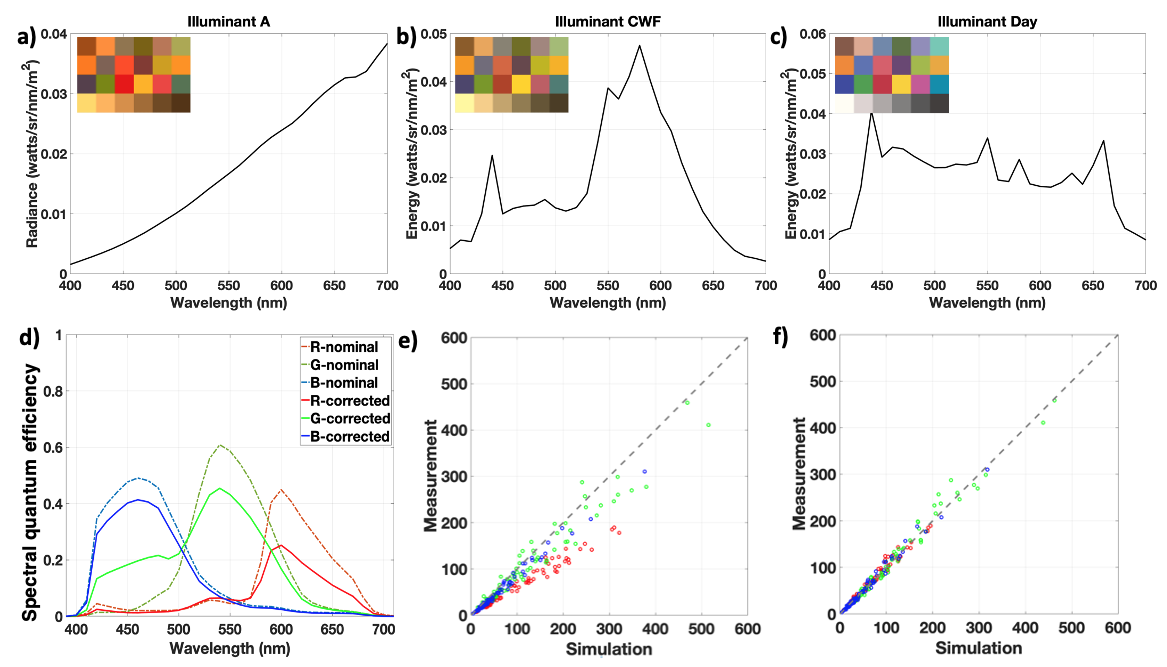}
  \caption{Sensor spectral quantum efficiency (QE). (a-c) Spectral energy of the lights used to illuminate the MCC.  d) Published (dashed lines) and transformed (solid lines) spectral QE for the Sony IMX363 sensor. e) Scatter plot comparing the predicted and simulated mean RGB values for 24 color patches based on the published spectral QE.  f) Scatter plot comparing predicted and simulated Mean RGB values based on the transformed spectral QE curves.}
  \label{Fig3}
\end{figure}

The dashed curves in Figure~\ref{Fig3}b show the spectral quantum efficiency (QE) of the Google Pixel 4a based on published sensor datasheets. Figure~\ref{Fig3}c shows the measured RGB pixel values for each of the 24 color patches against the RGB pixel values predicted by the product of the original sensor QE and the measured spectral radiance. The published  spectral quantum efficiency of the  Sony IMX363 does not include certain effects, such as optical and electrical crosstalk, or the channel gains. We calculated a 3x3 matrix, $M$, that transforms the original sensor QE curves and brings the predicted and measured RGB pixel values into better correspondence (see Figure~\ref{Fig3}f):
\begin{ceqn}
\begin{align}
\label{eq:ist}
[r^\prime, g^\prime, b^\prime] = [r,g,b]^TM
\end{align}
\end{ceqn}
where $r$, $g$, and $b$ are the vectors describing the published spectral quantum efficiency (QE) for red, green and blue channels (dashed lines, Figure~\ref{Fig3}d). 
The terms $[r^\prime, g^\prime, b^\prime]$ are the transformed QE vectors. The matrix $M$ minimizes the least squared error between measured and simulated sensor data for the 24 MCC color patches under the three different lights: 

\begin{ceqn}
\[M=
\begin{bmatrix}
0.532 & 0 & 0\\
0.06 & 0.70 & 0\\
0 & 0.36 & 0.84
\end{bmatrix}\]
\end{ceqn}

The diagonal entries of $M$ are a gain change in each of the three channels; the off-diagonal entries represent crosstalk between the channels. The transformed spectral QE data are the solid curves in Figure~\ref{Fig3}c.  These spectral QE curves produce a better match with the measured RGB values (compare Figure~\ref{Fig3}ef).  Hence we used the transformed curves in the simulations.

\section{Validations}
\begin{figure}[!hb]
  \includegraphics[width=1\columnwidth]{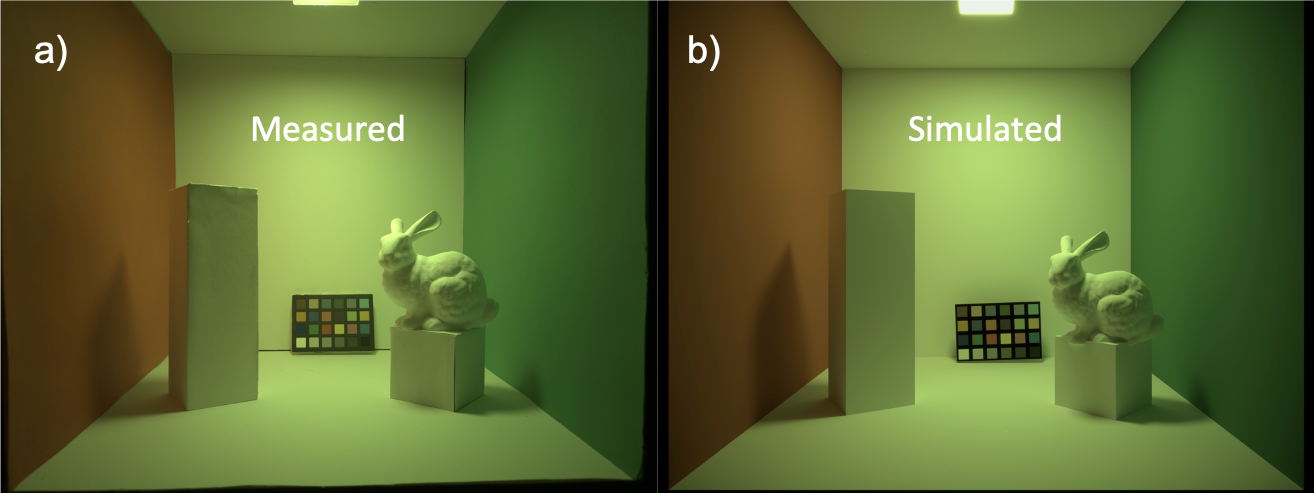}
  \caption{An example of measured and simulated images.}
  \label{Fig4}
\end{figure}

Figure~\ref{Fig4} shows real and simulated raw Google Pixel 4a camera images of a MCC target illuminated in the Cornell Box. The sensor data are formatted as Bayer color filter arrays. To simplify the comparison, we demosaicked the measured and simulated raw sensor image data using a bilinear interpolation.  

\subsection{Qualitative comparison}
The simulated and measured images share many similarities. Both represent the high dynamic range of the scene: The pixels in the light source area on the top are saturated while the shadows are simulated next to the cubes. Both camera images illustrate the effect of light inter-reflections: The side surface of the cube on the left is red due to the reflection from the left wall.

We do not expect that the measured and simulated sensor data can match on a pixel-by-pixel basis.  Rather, we only expect that critical properties of the two sensor data sets, including blur, chromatic properties, and noise, will match. We quantify the validity of the simulations by comparing small regions in the measured sensor data with the simulated sensor data. 

\subsection{Sensor digital values}
\begin{figure*}[!ht]
  \includegraphics[width=2\columnwidth]{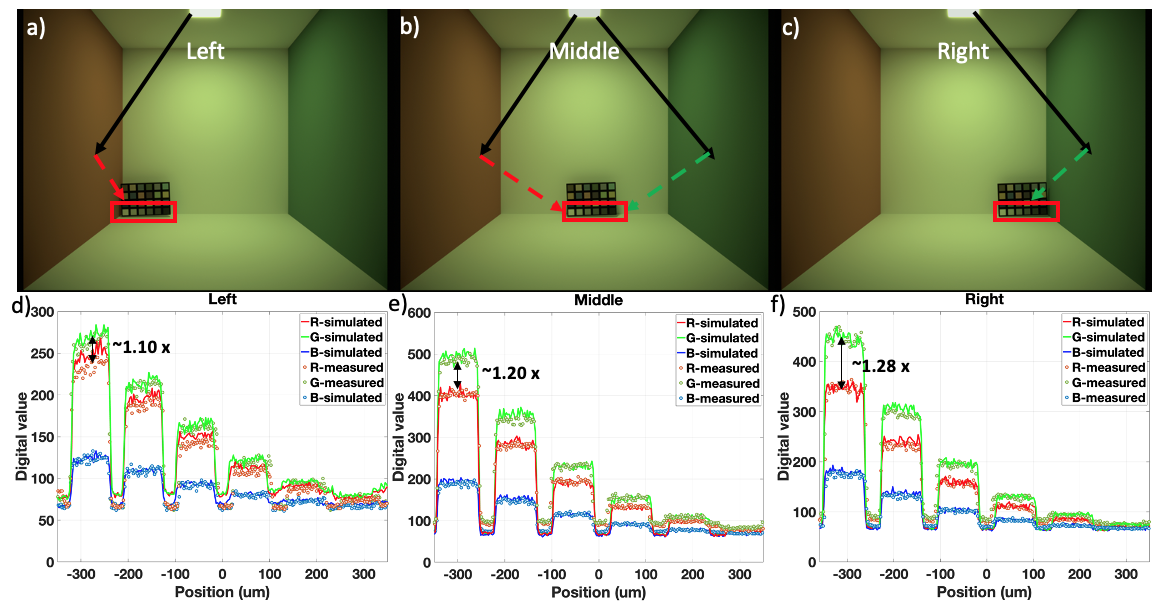}
  \caption{Analysis of sensor digital values. a-c): Simulated sensor images showing the MCC in different positions within the Cornell box. The arrows indicate the strong indirect lighting from the two colored walls. The red boxes outline the gray series of the MCC. d-f) The digital values from a line through the gray series of the simulated sensor data (solid) and measured sensor data (dots). The mean luminance of the simulated image was 21.5 cd/m2.  To match the levels of the simulated and measured data on the left and right, we multiplied the simulated data by 0.95 (right) and 0.75 (left).  We believe the scaling is necessary because we did not match the orientation of the MCC as we repositioned it against the back wall.}
  \label{Fig5}
\end{figure*}

The data in Figure \ref{Fig3}f show that the sensor simulations predict the mean RGB digital values for the 24 color patches in the MCC target illuminated by 3 different lights. Here, we evaluate the accuracy of the digital values when the MCC color target is placed in a more complex environment that includes surface inter-reflections. The simulated images in Figure~\ref{Fig5}a-c show the Cornell box with an MCC target placed at three different positions. We measured images that are very similar to these three simulated images. In both cases, we expect that the digital values measured from the MCC will vary as its position changes from near the red wall (left) to near the green wall (right). 

The graphs (Figures \ref{Fig5}d-f) show the RGB digital values for a horizontal line of image pixels through the bottom row (gray series) of the MCC. The solid lines show the simulated channels and the individual points show corresponding measurements. When the MCC is near the red wall the R/G ratio is highest, decreasing as the MCC position shifts towards the green wall.

The lowest pixel values - between the patches - are slightly higher for the measured sensor image than the predicted camera image.  This difference occurs because the reflectance of the black margins separating the color patches of the MCC is higher than the simulated reflectance (which we set to zero). Within each of the patches, the levels and the variance of the digital values are similar when comparing the measured and simulated data. The agreement between these two data sets confirms that the sensor noise model is approximately accurate. We explore the noise characteristics quantitatively in the next section.

\subsection{Sensor noise}
Sensor pixel values start from a noise floor, rise linearly with light intensity over a significant range, and then saturate as the pixel well fills up.  The signal measured at each pixel includes both photon and electrical noise.  In addition, there are variations between pixels in their noise floor levels (DSNU) and photoresponse gain (PRNU). These noise sources make different contributions depending on the mean signal level, and the combined effects of the noise sources contribute to the standard deviation of the digital values. 

\begin{figure}[!hb]
  \includegraphics[width=1\columnwidth]{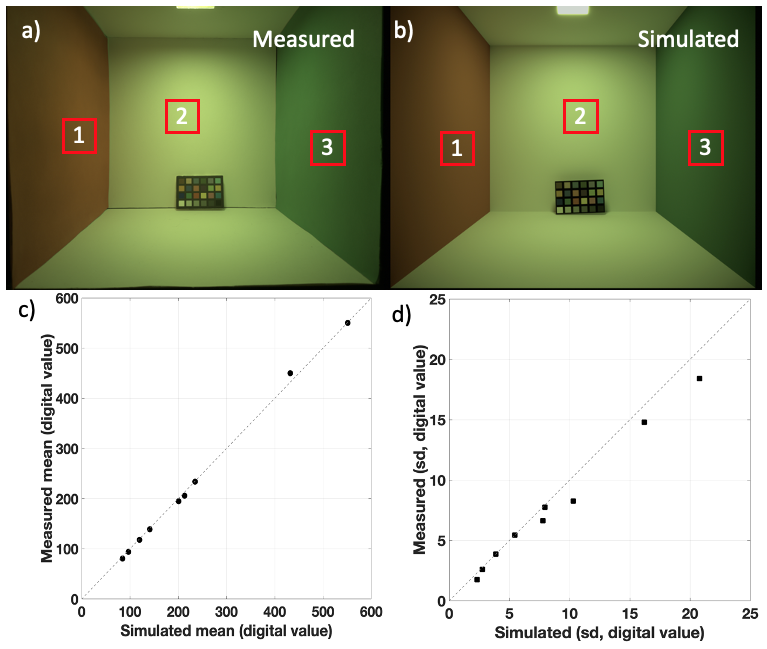}
  \caption{Sensor noise analysis. (a,b) We selected three regions in the measured and simulated images (red squares) where the mean RGB values of the measured and simulated sensor data were nearly equal. (c) The mean digital values for the nine values (RGB, 3 locations) are shown as a scatter plot.  (d) The standard deviations of the simulated and measured digital values shown as a scatter plot.}
  \label{Fig6}
\end{figure}

We selected three corresponding regions from the measured and simulated sensor images where the mean values match (Figure \ref{Fig6}ab), and compared the standard deviations in the measured and simulated digital values. Figure \ref{Fig6}c confirms that the mean levels in the three selected regions match closely. Figure \ref{Fig6}d shows that the standard deviations of the three channels in each of the three regions are similar. At the lower mean signal levels, the standard deviations are equal. At the higher mean levels, the simulated noise is slightly larger than the measured noise. The largest difference between the observed and expected standard deviation is about 2 digital values.

\section{Discussion}
The simulated images are similar to the sensor data measured from these moderately complex scenes (Figure~\ref{Fig4}). The simulation accounts for factors including object shape, surface reflectance, lighting geometry and spectral power distribution, interreflections, sensor spectral quantum efficiency, electronic noise and optics. The simulations do not predict a point-by-point match, but the mean signal and noise in corresponding regions agree to an accuracy of a few percent. This level of agreement is similar to the agreement between two cameras of the same model.

There are several limitations of this work. The validations are mainly confined to paraxial measurements and relatively uniform regions over the image. We plan additional evaluations to predict data at large field heights, more extensive depth of field, and higher dynamic range. There are at least two reasons why we do not think it will be possible to achieve pixel-by-pixel matches between the simulated and measured data: There will be minor differences between the real materials and the simulations, and some geometric differences will remain because we cannot perfectly match the camera position between simulation and reality. Finally, because the design of lenses used in Pixel 4a was unknown to us, we approximated the optics model with a shift-invariant wavefront model. When the design is known this approximation is not necessary. We are developing methods to lift this limitation through the use of black-box optics models.

There are many benefits of end-to-end image system simulations. First, simulations can be performed prior to selecting system components, making simulations a useful guide in system design. Second, simulations enable an assessment of two designs using the same, highly controlled, and yet complex input scenes.  Repeatable evaluation with complex scenes can be nearly impossible to achieve with real hardware components. Third, it is possible to use simulation methods to create large data sets suitable for machine learning applications. In this way, it is possible to understand whether hardware changes will have a significant impact on the performance of systems intended for driving, robotics and related machine learning applications.

We are using end-to-end image systems simulations methods for a wide range of applications.  These include designing and evaluating novel sensor designs for  high dynamic range imaging \cite{jiang2017learning}, underwater imaging \cite{Blasinski2017-qb}, autonomous driving \cite{Liu2021-mr} and fluorescence imaging \cite{Farrell2020-tb, Lyu2021-ae}.  End-to-end image systems simulations can accelerate innovation by reducing many of the time-consuming and expensive steps in designing, building and evaluating image systems. We view the present work as a part of an ongoing process of validating these simulations.

\section{Acknowledgements}
We thank Guangxun Liao, Bonnie Tseng, Ricardo Motta, David Cardinal, Zhenyi Liu, Henryk Blasinski and Thomas Goossens for useful discussions.

\section{Reproducibility}
The suite of software tools (ISET3D, PBRT and ISETCam) are open-source and freely available on GitHub (see https://github.com/ISET).  Scripts to reproduce the figures in this abstract are in that repository.


\small
\bibliographystyle{ieeetr}
\bibliography{EICornellBox}



\begin{biography}
Zheng Lyu is a PhD candidate in the Department of Electrical Engineering at Stanford University. 

Krithin Kripakaran, De Anza College and a research assistant at Stanford University.

Max Furth, a student at Brighton College, England.

Eric Tang, a student at Gunn High School, Palo Alto, CA.  

Brian A. Wandell is the Isaac and Madeline Stein Family Professor in the Stanford Psychology Department and a faculty member, by courtesy, of Electrical Engineering, Ophthalmology, and the Graduate School of Education. 

Joyce Farrell is the Executive Director of the Stanford Center for Image Systems Engineering and a Senior Research Associate in the Department of Electrical Engineering at Stanford University. 

\end{biography}

\newpage
\section{Appendix}

\begin{table}[!h]
\caption{Table1. Sony IMX363 sensor specification.}
\label{tab:sensor}
\begin{center}       
\begin{tabular}{|l|p{0.30\columnwidth}|p{0.35\columnwidth}|} 
\hline
\textbf{Properties} & \textbf{Parameters} & \textbf{Values(units)} \\ \hline\hline
\multirow{2}{*} {Geometric} & Pixel Size & [1.4, 1.4] ($\mu$m) \\
                            & Fill Factor & 100 ($\%$) \\ \hline
\multirow{6}{*} {Electronics} & Well Capacity & 6000 ($e^-$) \\
                         & Voltage Swing & 0.4591 (volts) \\
                         & Conversion Gain & $ 7.65 \times 10^{-5} (volts/e^-$) \\
                         & Analog Gain & 1 \\
                         & Analog Offset & 0 (mV) \\
                         & Quantization Method & 12 (bit) \\ \hline      
\multirow{4}{*} {\makecell[l]{Noise Sources \\ @Analog gain=1}} & DSNU & 0.64 (mV) \\
                          & PRNU & 0.7 (\%) \\
                          & Dark Voltage & 0 (mV/sec) \\
                          & Read Noise & 5 (mV) \\ \hline                         

\end{tabular}
\end{center}
\end{table} 

\end{document}